\documentclass[preprint,showpacs,amsmath,amssymb]{revtex4}
\usepackage{mathrsfs}

\usepackage{graphicx,color}
\usepackage{dcolumn}
\usepackage{bm}
\def\slash#1{{#1\!\!\!/}}
\newcommand{\nb}{\nonumber}
\begin{document}


\title{Probing the charm-Higgs Yukawa coupling via Higgs boson decay to $h_c$ plus a photon}

\author{Song Mao$^a$}
\author{Yang Guo-He$^a$}
\author{Li Gang$^a$}\email{lig2008@mail.ustc.edu.cn}
\author{Zhang Yu$^b$}
\author{Guo Jian-You$^a$}

\affiliation{$^a$ School of Physics and Material Science, Anhui University, Hefei, Anhui 230039, P.R.China}
\affiliation{$^a$ Institute of Physical Science and Information Technology, Anhui University, Hefei, Anhui 230039, P.R.China}

\date{\today}

\begin{abstract}
In this paper, we investigate the decay of Higgs boson to $h_c$ plus a photon in the NRQCD theoretical framework.
Comparing with the Higgs decay to $J/\psi$ plus a photon channel, this process has not indirect contribution,
can be used to detect the Yukawa coupling of Higgs and charm quarks. The results show that the decay branch ratio of this process is about $10^{-8}$. If we takes into account the $10^{-3}$ efficiency in the $h_c$ detection, no events will be available even in the case of $30ab^{-1}$ luminosity at FCC-pp with 100 TeV center of mass energy. However, if the detection efficiency of $h_c$ is greatly improved in the future, this process will play an important role at linear $e^+e^-$ future colliders and at LHCb. Moreover, this process should be also play an important role when the anomalous charm Yukawa couplings are larger and direct sensitivity.
\end{abstract}

 \maketitle

\section{Introduction}
\par
The discovery of the Higgs boson is a triumph of the LHC \cite{higgs1,higgs2,higgs3,higgs4} and it is also a success for the
Standard Model (SM) with its minimal Higgs sector of electroweak (EW) symmetry breaking (EWSB).
After discovery of the Higgs boson, one of the most tasks is to determine its properties, such as its spin,
CP, width, and couplings. Up to now, all measurements of the Higgs boson properties are so far indicating that the
observations are compatible with SM Higgs. For the main Higgs discovery modes $\gamma\gamma$, $ZZ$ and $WW$, the couplings to gauge bosons are measured
directly, which are fixed through the well measured diboson decays of
the Higgs determined at the $20\sim30\%$ level. Direct evidence for the Yukawa coupling of the Higgs boson to the top \cite{Aaboud:2017jvq}
and bottom \cite{Aaboud:2017xsd,Sirunyan:2017elk} quarks was recently obtained. Measurements of the Yukawa coupling of the Higgs
boson to the first- and second-generation quarks are need to do in the near future.

\par
The Standard Model Higgs boson direct decaying to a pair of charm quarks, through associated production of
the Higgs and Z bosons, in the decay mode $HZ \to l^+l^-c \bar{c}$ is studied by the ATLAS experiment at the LHC in Ref.\cite{Aaboud:2018fhh}.
Charm jets are particularly challenging to tag because c-hadrons have shorter lifetimes and decay to fewer charged particles than b-hadrons.
The largest uncertainty is due to the normalization of the dominant Z$+$jets background. Therefore, the charm quark Yukawa coupling are hard to accurate measurement in hadron colliders through the direct $H \to c\bar{c}$ decay, owing to large QCD backgrounds, and challenges in jet flavor identification \cite{Perez:2015aoa,Perez:2015lra}.

\par
Heavy quarkonium $J/\psi$ is a $c\bar{c}$ bound state and can decay to $e^+e^-$ or $\mu^+\mu^-$. These leptonic decay modes are clean channels in experiments and suppress large QCD backgrounds. In Ref.\cite{Bodwin:2013gca}, The authors showed that the exclusive decays of the Higgs boson to vector mesons can probe the Yukawa couplings of first- and second-generation quarks and serve as searching New Physics (NP) beyond SM at future runs of the LHC. Then, Higgs rare decay to a vector quarkonium ($J/\psi,\Upsilon$) received considerable attention \cite{Kagan:2014ila,Gao:2014xlv,Modak:2014ywa,Delaunay:2013pja}. The relativistic correction for Higgs boson decay to an S-wave vector
quarkonium plus a photon have been calculated in Ref. \cite{Bodwin:2014bpa}.
In Ref. \cite{Chao:2016usd}, the authors evaluated the NLO corrections to
$H \to J/\psi + \gamma$ and find that the direct contribution are greatly reduced by the NLO QCD correction.
A search for Higgs and Z bosons decaying to $J/\psi$ and $\Upsilon$ is performed
in integrated luminosities $20.3 fb^{-1}$ with the ATLAS detector at 8 TeV LHC. No significant
excess of events is observed above expected backgrounds and $95\%$ CL upper limits are placed on the
branching fractions \cite{Aad:2015sda}.

\par
There is another problem needs to be solved via Higgs decay to $J/\psi \gamma$ to study the charm quark Yukawa couplings.
In this decay channel, there are two part contributions for total decay width: one come from the direct contribution, which is related to charm quark Yukawa  coupling, and the other part come from the indirect contribution, which arises from $H \to \gamma^\ast \gamma$ with
virtual $\gamma$ substantially converting into $J/\psi$, and not related to charm quark Yukawa coupling. However, the width of indirect decay is much larger than that of direct decay. Suppressing the indirect contribution becomes an important and unavoidable question. Previously, all the people focus on Higgs decay to S state charmonium $J/\psi$. In fact, if the P state charmonium is selected as a candidate, like $h_c$, which has quantum
numbers  $J^{PC} = 1^{+-}$, due to the CP invariance of Quantum Electrodynamics (QED), virtual $\gamma$ converting into $h_c$ is forbidden, the contribution of the indirect decay can be completely removed \cite{Fleming:1998md}.

\par
$h_c$ meson is the lowest spin-singlet P-wave charmonium, which is first found via the process $p\bar{p} \to h_c \to J/\psi \pi^0$ at Fermilab E760 experiment in 1992~\cite{Armstrong:1992ae}. Then, the $h_c$ state is measured by Fermilab E835, CLEO-c, BESIII experiments~\cite{Andreotti:2005vu,Rosner:2005ry,Rubin:2005px,Ablikim:2010rc,Ablikim:2012ur}. Its C-parity was established by radiative decay $h_c \to \eta_c \gamma$ \cite{Ablikim:2010rc}. In recent years, the production and decay of $h_c$  has been studied at $e^+e^-$ and hadron colliders \cite{Qiao:2009zg,Jia:2012qx,Wang:2012tz,Wang:2014vsa,Sun:2018yam}. In this paper, we will investigate the process
$H \to h_c + \gamma$ within the non-relativistic QCD (NRQCD) framework by applying the covariant projection method \cite{p2}.

\par
The paper is organized as follows: we present the details of the calculation strategies in Sec.II.
The numerical results are given in Sec.III. Finally, a short summary and discussions
are given.

\section{Calculation descriptions}

\par
\begin{figure}[!htb]
\begin{center}
\begin{tabular}{cc}
{\includegraphics[width=0.8\textwidth]{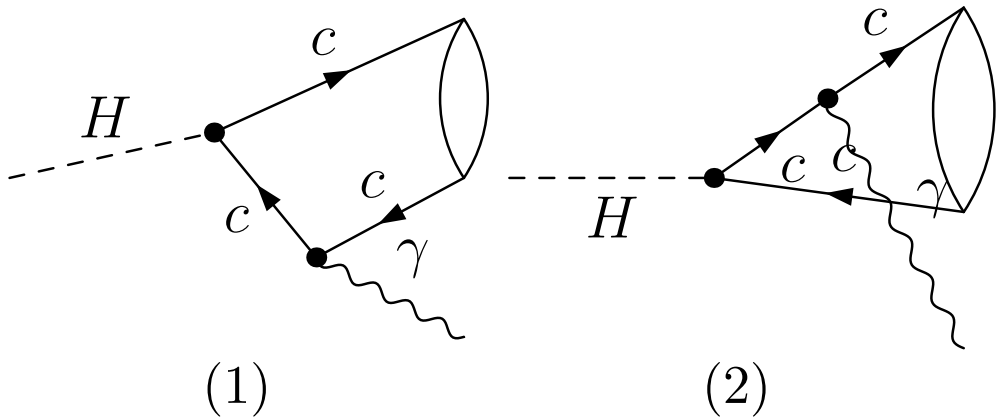}}
\end{tabular}
\end{center}
 \vspace*{-0.7cm}\caption{The Feynman diagrams for the $H\to c\bar{c}[^1P_1^{(1)}]+ \gamma $ decay process at the LO} \label{f1}
\end{figure}

In this section, we present the calculation for the decay process $H \to h_c + \gamma$.
There are two Feynman diagrams for $H \to h_c + \gamma$ at parton level in leading order(LO), which are
drawn in Fig.\ref{f1}. We calculate the amplitudes by making
use of the standard methods of NRQCD factorization \cite{bbl}, which provides a rigorous theoretical framework for the description of heavy quarkonium production and decay. In the NRQCD, the idea of perturbative factorization is applied, the process of production and decay of heavy quarkonium is separated into two parts: short distance part, which allows the intermediate $Q\bar{Q}$ pair with quantum numbers different from those of the physical quarkonium state, and the long distance matrix elements (LDMEs), which can be extracted from experiments. NRQCD is an effective factorization method and has became the standard tool for theoretical calculations for heavy quarkonium \cite{Brambilla:2010cs}. The
partonic process $H\to c\bar{c}+ \gamma $ at LO is denoted as:
\begin{equation}\label{emission}
H(p_1)\to c(p_2)\bar{c}(p_3)+\gamma(p_4).
\end{equation}
The amplitudes for these two diagrams are given by
\begin{eqnarray}
{i\cal M}_{i1} =  \bar{u}_{si}(p_2)\cdot\frac{-i e m_c}{2m_W s_W}\cdot \frac{i}{\rlap /p_1-\rlap /p_2-m_c}\cdot i\frac{2}{3}e \gamma^\mu \cdot v_{s'j}(p_3)
                \epsilon_{\mu}^*(p_4),~ \nonumber \\
{i\cal M}_{i2} = \bar{u}_{si}(p_2) \cdot i\frac{2}{3}e \gamma^\mu \cdot \frac{i}{\rlap /p_1-\rlap /p_3-m_c} \cdot\frac{-i e m_c}{2m_W s_W}\cdot v_{s'j}(p_3)
                \epsilon_{\mu}^*(p_4).~ \nonumber \\
\end{eqnarray}

where s and $s'$ are spin indices, i and j are color indices of the outgoing c quark and ${\bar c}$ quark, respectively.
The relative momentum between the $c$ and $\bar{c}$ is defined as $q
= (p_2 - p_3)/2$, and the total momentum of the bound state $c\bar{c}$ is defined
as $p = p_2 + p_3$. Then, we obtain the following relations among
the momenta:

\begin{eqnarray}
p_{2} &=& \frac{1}{2}p + q, \;\;\; p_{3}= \frac{1}{2}p - q,  \;\;\; p \cdot q = 0,\nonumber \\
p^{2}_{2} &=& p^{2}_{3}=m^{2}_{c}, \;\;\; p^2 = E^2,\;\;\; q^2 =
m_c^2-E^2/4=-m_c^2 v^2. \label{eq:momdefs}
\end{eqnarray}

In the $c\bar{c}$ rest frame, $p = (E, 0)$ and $q = (0, q)$. In the non-relativistic $v=0$ limit, $p^2 = 4 m^2_c,~~ q^2 = 0$. In
order to produce a $h_c$, the $c\bar{c}$ pair must be produced in
a spin-singlet, color-singlet fock state with orbital angular momentum $L=1$.
The short-distance amplitudes are obtained by differentiating the spin-singlet, colour-singlet
projected amplitudes with respect to the momentum $q$ of the heavy quark in the $c\bar{c}$ rest frame, and then
setting relative momentum $q$ to zero. As following the notations in Ref.\cite{p2}, the short-distance amplitudes are expressed as:

\begin{eqnarray}
{\cal M}_{^1P_0^{(1)} } = {\epsilon}_{\beta}
\frac{d}{q_\beta} {\rm Tr}\left.\left[{\cal C}_{1}\,
\Pi_{0}
{\cal M}\right]\right|_{q=0}, \nonumber
\end{eqnarray}

where ${\epsilon}_{\beta}$ is the polarization vector of $^1P_1^{(1)}$ state, and the spin-singlet projector is given by
\begin{eqnarray}
&&{\Pi}_{0}
     = {1\over{\sqrt{8m_c^3}}} \left({{\slash{p}}\over 2} - \slash{q} - m_c\right)
       \gamma_5 \left({{\slash{p}}\over 2} + \slash{q} + m_c\right)\, .
\end{eqnarray}

The colour singlet state will be projected out with the following operator:
\begin{eqnarray}
&&{\cal C}_1 = {{\delta_{ij}}\over{\sqrt{N_c}}}
\label{proj_sing}
\end{eqnarray}
The amplitude ${\cal M}$ is obtained by truncating the external spinor $\bar{u}(p_2)$ and $v(p_3)$ in Fig.\ref{f1}.
The trace is sum over all the Lorenz and colour indices. The selection of the appropriate
total angular momentum quantum number is done by performing the proper polarization sum. Here, we define:
\begin{equation}
\Pi_{\alpha\beta} \equiv -g_{\alpha\beta} + {{p_\alpha p_\beta}\over{M^2}}
\; ,
\end{equation}
where $M=2m_c$.

After the application of this set of rules, the short-distance contribution to the differential decay width
for $H\to c\bar{c}[^1P_1^{(1)}]+ \gamma $ process reads:
\begin{eqnarray}
&&d\hat\Gamma(H \to c\bar{c}[^1P_1^{(1)}]+ \gamma) = {1\over{32 \pi^2}}
     |{\cal M}_{^1P_1^{(1)}}|^2\;{|p|\over{m_H^2}}d \Omega \, ,
\end{eqnarray}
where $|p|=\frac{m_H^2-m^2_{h_c}}{2 m_H}$  and $m_H$ represent the Higgs boson
mass. $d\Omega=d\phi d(cos\theta)$ is the solid angle of particle $h_c$.

\begin{eqnarray}
&&{|{\cal M}_{^1P_1^{(1)}}|^2 = {{256 \pi^2 \alpha^2 \overline{m}_c(\mu)^2}\over{3 m_c^3 m_W^2s_W^2}}}
\label{proj2}
\end{eqnarray}

where $\overline{m}_c(\mu)$ appeared in charm quark Yukawa coupling is the running mass of charm quark \cite{Djouadi:2005gi}.
In the modified minimal subtraction or $\overline{MS}$ scheme, the relation between the pole masses
and the running masses at the scale of the pole mass, $\overline{m}_c(m_c)$, can be expressed as

\begin{eqnarray}
\overline{m}_c(m_c) &=& m_c[1-\frac{4}{3}\frac{\alpha_s(m_c)}{\pi}+(1.0414N_f-14.3323)\frac{\alpha^2_s(m_c)}{\pi^2} \nonumber \\
                    &+& (-0.65269N^2_f+26.9239N_f-198.7068)\frac{\alpha^3_s(m_c)}{\pi^2})]
\end{eqnarray}

where $\alpha_s$ is the $\overline{MS}$ strong coupling constant evaluated at the scale of the pole mass $\mu = m_c$.
The evolution of $\overline{m}_c$ from $m_c$ upward to a renormalization scale $\mu$ is
\begin{eqnarray}
\overline{m}_c(\mu) = \overline{m}_c(m_c)\frac{c[\alpha_s(\mu)/\pi]}{c[\alpha_s(m_c)/\pi]}
\end{eqnarray}
with the function c, up to three-loop order, given by
\begin{eqnarray}
c(x) &=& (25x/6)^{12/25} [1 + 1.014x + 1.389 x^2 + 1.091 x^3]~for~m_c < \mu < m_b \nonumber \\
c(x) &=& (23x/6)^{12/23} [1 + 1.175x + 1.501 x^2 + 0.1725 x^3]~for~m_b < \mu < m_t \nonumber \\
c(x) &=& (7x/2)^{4/7}[1 + 1.398x + 1.793 x^2 - 0.6834 x^3]~for~m_t < \mu
\end{eqnarray}

Then, the total decay width is

\begin{eqnarray}
\Gamma (H\to h_c+\gamma) =  \hat{\Gamma}(H\to c\bar{c}(^1P_1^{(1)})+\gamma) \frac{< {\cal{O}}^{h_c}(^1P_1^{(1)})>}{2 N_c N_{col} N_{pol}},
\end{eqnarray}
where $N_{col}$ and $N_{pol}$ refer to the number of colors and polarization states
of the  $c\bar{c}$ pair produced.  The color-singlet
states $N_{col}=1$, and $N_{pol}=3$ for polarization vector in 4 dimensions.
$< {\cal{O}}^{h_c}(^1P_1^{(1)})>$ is the vacuum expectation value of the operator ${\cal{O}}^{h_c}(^1P_1^{(1)})$,
$2 N_c$ is due to the difference between the conventions in Ref.~\cite{p2} and Ref.~\cite{bbl}.

\section{Numerical results and discussion}
\par
In this section, we present our numerical results for the $H\to h_c+\gamma$ decay.
The relevant input parameters are set as follows \cite{pdg}:
\begin{eqnarray}
&\alpha^{-1}&=137.036,~m_H = 125.09~{\rm GeV},~m_Z = 91.1876~{\rm GeV},~ m_W = 80.385~{\rm GeV}, \\
&m_c&=1.64~{\rm GeV},~s_W^2 = 1 - m_W^2/m_Z^2,~m_{h_c} = 3.76~{\rm GeV},
\end{eqnarray}

The LDME of $<{\cal O}^{h_c}[^1P_1^{(1)}]> $ can be expressed in terms of radial derivative of the wave function
of quarkonium at the origin $<{\cal O}^{h_c}[^1P_1^{(1)}]>  =  \frac{27}{2 \pi}|R'_P(0)|^2$, where $|R'_P(0)|^2=0.075GeV^5$
from the potential model calculations has been used in our calculation.
The value of color-singlet Long Distance Matrix Elements (LDME) is set as
$<{\cal O}^{h_c}[^1P_1^{(1)}]>  = 0.32~{\rm GeV}^5$ \cite{Eichten:1995ch,Chao:2016chm,Wang:2014vsa}.

Since the mass of charm quark in the charm Yukawa coupling is dependent on the renormalization scale, the strength of the charm Yukawa coupling is also dependent on the renormalization scale. We take the Higgs mass as the central value of the renormalization scale for processes $H\to h_c+\gamma$, and the short distance theoretical uncertainty is estimated by the renormalization scale range from 1/2$m_H$ to 2$m_H$. In Table.\ref{tab:mu}, we list the running charm mass and decay width at different renormalization scales, where $m_c$ is taken as the pole mass 1.64 GeV except charm mass in the charm Yukawa coupling, $m_{h_c}$ is taken as 3.76 GeV.

\begin{table}[t]
  \centering
  \begin{tabular}{|c|ccc|}
  \hline\hline
$\mu$ & $M_H/2$ & $M_H$ & $2 M_H$  \\
\hline
$\overline{m}_c(\mu)(GeV)$ & 0.66&0.62&0.51 \\
${\Gamma}(H\to h_c+\gamma)\times10^{11}$ & 0.86&0.76&0.51 \\
\hline\hline
\end{tabular}
\caption{The renormalization scale $\mu$ dependence of the decay widths for the process $H\to h_c+\gamma$} \label{tab:mu}
\end{table}

The total width of a 125 ${\rm GeV}$ SM Higgs boson is $\Gamma(H) = 4.07\times 10^{-3} {\rm GeV} $, with a relative
uncertainty of $^{+4.0\%}_{-3.9\%}$\cite{pdg,deFlorian:2016spz}. Using this width of Higgs boson decays, we obtain the following results for the branching
fraction in the SM:
\begin{eqnarray}
{\cal B}(H\to h_c+\gamma) =0.187\times10^{-8}~
\end{eqnarray}.

If there is new physics beyond the Standard Model(SM), charm Yukawa coupling strength may be different from that in SM.
In order to consider the theoretical uncertainty, we assume that the Yukawa coupling strength of charm quark and Higgs boson is deviation from the coupling in SM. The deviations from the SM are implemented as scale factors ($\kappa^2$) of Higgs couplings relative to their SM
values, and it is defined as:
\begin{eqnarray}
g_{Hc\bar{c}} = \kappa \cdot g_{Hc\bar{c}}^{SM}
\end{eqnarray}
such that $\kappa = 1$ in SM. In Table.\ref{tab:kappa}, we illustrate the parmeter $\kappa^2$ dependence of the decay widths for the process $H\to h_c+\gamma$. The mass of $h_c$ is set as 3.76 GeV, $m_c = 1.64~GeV$, the renormalization scale is set as $\mu = m_H$. When $\kappa^2$ running from 0.1 to 10, the decay widths vary from $0.076\times10^{-11}{\rm GeV}$ to $7.6\times10^{-11}{\rm GeV}$ for the processes $H\to h_c+\gamma$ , respectively.

\begin{table}[t]
  \centering
  \begin{tabular}{|c|ccccccc|}
  \hline\hline
$\kappa^2$ & 0.1 & 0.2 & 0.5 & 1 & 2 & 5 & 10 \\
\hline
${\Gamma}(H\to h_c+\gamma)\times10^{11}$ & 0.076&0.152&0.38&0.76&1.52&3.8&7.6 \\
\hline\hline
\end{tabular}
\caption{The parmeter $\kappa^2$ dependence of the decay widths for the process $H\to h_c+\gamma$} \label{tab:kappa}
\end{table}

In experiment, $h_c$ is detected mainly by the following three decay channels:
\begin{eqnarray}
h_c &\to& \pi_0 J/\psi \to l^+l^- \gamma\gamma, \nb \\
h_c &\to& \eta_c \gamma \to p \bar{p}\gamma, \nb \\
h_c &\to& \eta_c \gamma \to \gamma \gamma \gamma.
\end{eqnarray}
The branching ratio of $h_c \to \pi_0 J/\psi$ and $h_c \to \eta_c \gamma$ are estimated to be about 0.5\% \cite{b8} and 50\%
\cite{b91,b92,b93,b94} in theory, respectively. In these decay chains, the branching ratio of $J/\psi$ decaying into $l^+l^-$ is about $12\%$ \cite{pdg}, $\pi^0$ almost completely decays into $\gamma \gamma$, the branching ratio of $\eta_c$ decaying into $p\bar{p}$ is about $0.13\%$, and into $\gamma \gamma$ with a ratio of $0.024\%$ \cite{pdg}.
The total cross section of Higgs production at 14TeV LHC is about 62 pb. If the integral luminosity of LHC reaches 3000 $fb^{-1}$,
it will accumulate about $2\times10^8$ Higgs events. Considering the decay branching ratio of $H\to h_c +\gamma$, there will be about 0.4 events of $h_c +\gamma$ decaying from Higgs boson. If we takes into account the $10^{-3}$ efficiency in the $h_c$ detection in hadron colliders, no events will
be available even in the case of $30ab^{-1}$ luminosity at FCC-pp with 100 TeV center of mass energy. However, if the detection efficiency of $h_c$ is greatly improved in the future, this process will play an important role at linear $e^+e^-$ future colliders and at LHCb. Moreover, like $h \to J/\psi + \gamma$, this process should be also play an important role when the anomalous (large $\kappa$) charm Yukawa couplings are larger and direct sensitivity.

\section{Summary}
Compared to the process of Higgs decay to $J/\psi$ plus a photon, the process of Higgs decaying to $h_c$ plus a photon can greatly reduce the indirect contribution and can be used to directly detect the coupling of Higgs and charm quarks.
In this paper, we calculated the decay width and decay branch ratio of Higgs decay to $h_c$ plus a photon in the theoretical framework of NRQCD.
We found that the branch ratio is about $0.187\times 10^{-8}$, and there will be no enough events to produce with integrated luminosity $3000fb^{-1}$ at the 14 TeV LHC. If the detection efficiency of $h_c$ is taken into consideration, it is difficult to observe it on the LHC.
However, if the detection efficiency of $h_c$ is greatly improved in the future, this process will play an important role at linear $e^+e^-$ future colliders and at LHCb and it also will play an important role when the anomalous (large $\kappa$) charm Yukawa couplings are larger and direct sensitivity.

\section{Acknowledgments}
This work was supported in part by the National Natural Science Foundation of China (No.11305001, No.11575002, No.11805001) and the Key Research Foundation of Education Ministry of Anhui Province of China under Grant (No.KJ2017A032).


\end{document}